# Upper-Level Physics Students' Perceptions of Physicists

Paul W. Irving and Eleanor C. Sayre, Kansas State University, 116 Cardwell Hall, Manhattan, KS 66506
Email: paul.w.irving@gmail.com, esayre@ksu.edu

**Abstract:** As part of a longitudinal study into identity development in upper-level physics students, we used a phenomenographic research method to examine students' perceptions of what it means to be a physicist. The results revealed four different categories. We find a clear distinction in the exclusivity students associate with being a physicist and the differences in the importance of research and its association with being a physicist. A relationship between perceptions of physicists and goal orientation is indicated.

## Introduction

The development of a professional identity is a key part of student development. The development of an appropriate subject specific identity is a strong influence on retention within a discipline. In the context of physics there is a strong link between level of identification with being a physicist and whether or not a student has settled on a physical science career (Barton & Yang, 2000). Understanding how students get to the point where they have achieved a strong level of identification with being a physicist is a complex undertaking and involves investigating multiple aspects of students' relationship with physics. These aspects include: examining students as they practice (do physics); who they are practicing with (their community of practice (Lave & Wenger, 1991)) and their personal identity (experiences, attributes and perceptions that shape their relationship with physics). In this poster we explore upper-level physics students' perceptions of what it means to be a physicist.

## Context and Methodology

The primary data sources are semi-structured phenomenographic interviews with students recruited from upper-level physics courses at a large midwestern university. Phenomenographic research typically focuses on a relatively small number of subjects and identifies a limited number of qualitatively different and interrelated ways in which the subjects experience and perceive a phenomenon. Our interviews focused on several topics including identity formation, epistemological sophistication, and metacognition. 21 students ranging from sophomores to seniors (18 male) participated. Our approach, based on the theory of variation and awareness (Marton & Booth, 1997) is grounded in the assumption that there are a limited number of qualitatively different ways in which something that is experienced and perceived – such as what it means to be a physicist – can be understood. From the theory, the goals of a phenomenographic analysis are to seek these qualitatively different ways and the variations thereon. In our study, the students' responses to questions were analyzed using an iterative process involving examinations of the video and transcripts with a particular focus of awareness on one aspect of the data with each sitting (for example: their experiences with research). The variation amongst critical aspects was utilized to form descriptions (outcome space) of the different perceptions of physicists. It is important to note that the interviews were focused on discovering the different students perceptions of being a physicist (phenomenographical) and how that relates to their current studies at a particular time in their undergraduate career as opposed the personal meaning of being a physicist to these students (phenomenological).

## Results

From our analysis of the interview data, four distinct categories of description emerged for students' perceptions of physicists. We briefly describe the categories below.

### Physicists Are Researchers Who Answer the Unanswered Questions. (High Research/Mastery):

Students in this category believe that physicists answer the unanswered questions and they do so by conducting "new" research. *"Abbey: I would say a physicist is someone who is trying to answer some of the unanswered questions, trying to prove the impossible."* Students have a passion for answering the questions that remain unanswered and see physics as a tool to do so. To do physics research is an important occupation but is dependent on the fact that one is not a physicist unless one is doing research. These students also discuss that they enjoy physics because it encourages a more complete understanding of a phenomenon. A focus on more complete understanding and the fact that these students do not believe that they will be physicists until they have contributed new knowledge/understanding to the physics community indicates a mastery orientation to physics.

### Physicists Are Scientists Who Practice and Are Knowledgeable of Physics (High Research/Performance)

Students in this category had a similar perception that doing research was important to be classified as a physicist. However, these students did not posit research as being a grand endeavor of answering unknown questions nor were they able to elaborate on what doing physics research might involve. Instead research is just "something" that is done to be "active" within the field. *"Dylan: Mostly the research I think that makes them a physicist, I mean, you get your degree and everything but its actually doing something that makes you a physicist."* These students when talking about the subject are orientated more towards performance. They believe that it is important to obtain a lot of knowledge of physics and that a certain amount of physics knowledge must be obtained to be considered a physicist.

### Physicists Are People with a Certain Mindset (Low Research/Mastery)

The students in this category believe themselves to be physicists already as they are involved in some capacity with physics, to them it is a way of approaching/viewing the world so that they understand it in a physical sense. *"Larry: I don't think you necessarily have to be, like, doing research actively to be a physicist, I just think you have to have an appreciation for physics and be involved with it in some capacity."* To these students there is no amount of knowledge to be obtained or no activity that must be completed to be a physicist besides having an inherent interest in the subject. The focus on understanding the world marks these students as being orientated towards mastering the subject.

### Physicists Are People Who Are Committed to Physics (Low Research/Performance)

Students in this category are focused on the fact that they are committed to the subject. The fact that someone has made a commitment to learning more about physics, is what makes someone a physicist. These students were also more likely to declare themselves a physicist at this point in their academic career as they have made a commitment to studying this subject for their extended future. *"Sally: I think they are a physicist when they have declared a commitment to it, (filled pause) to the subject, whether that is declaring a major or spending time studying it...but making a definite commitment to the subject."* These students place an emphasis on their knowledge being evaluated by an external source as an important element to being a physicist, which indicates that these students are orientated towards performance.

|  | | Research | |
|---|---|---|---|
|  | | High Importance Placed | Low Importance Placed |
| **Orientation** | Mastery | Researchers who answer unanswered questions<br>8 | People with a certain mindset<br>3 |
|  | Performance | Scientists who practice and knowledgeable of physics<br>7 | People who are committed to physics<br>3 |

Figure 1. Categories and number of students categorized into each.

The majority of students' emphasis on research (Figure 1.) is likely a product of the university being studied. The faculty transmit a consistent message that research and obtaining a research experience is important. This message is received but what research actually entails and why it is important for a student to be involved in it in some capacity is content of this message that is not received by a substantial portion of the students involved in our study.

## Conclusions

We discovered four perceptions of what it means to be a physicist and a found a relationship between these perceptions and goal orientation. This relationship should inform program developers that an emphasis on research should include an emphasis on what research entails. To many of the students in this study, research is just one more goal that one must achieve on ones trajectory to becoming a physicist. Perceiving research as a *goal to be achieved* rather than an *essential experience to be learned from* could have adverse effects to the development of a student's identity, their approach to research experiences, and ultimately their commitment to pursuing physics. This study is a snapshot of a particular time in this group of students undergraduate career and we expect answers to several of these questions to emerge through a longitudinal study of these students.